\documentclass[ams,amsmath,nofootinbib]{revtex4}
\usepackage{verbatim}					
\usepackage{graphicx}
\usepackage{anysize}
\usepackage{amsfonts}

\newcommand{\be}{\begin{equation}}
\newcommand{\ee}{\end{equation}}
\newcommand{\benn}{$$}
\newcommand{\eenn}{$$}
\newcommand{\bea}{\begin{eqnarray}}
\newcommand{\eea}{\end{eqnarray}}
\newcommand{\beann}{\begin{eqnarray*}}
\newcommand{\eeann}{\end{eqnarray*}}
\newcommand{\f}{\frac}

\let\a=\alpha \let\b=\beta  \let\g=\gamma  \let\d=\delta
   
\let\m=\mu    \let\n=\nu          \let\r=\rho 
\let\s=\sigma   
 \let\eps=\epsilon

\def\det{\mathop{\rm det}\nolimits}
\def\tr{\mathop{\rm Tr}\nolimits}
\def\arctanh{\mathop{\rm arctanh}\nolimits}
\def\arctan{\mathop{\rm arctan}\nolimits}
\def\ln{\mathop{\rm ln}\nolimits}

\DeclareMathAlphabet{\mathpzc}{OT1}{pzc}{m}{it}

\newcommand{\Ref}[1]{(\ref{#1})}

\begin{document}

\title{Scalar-Tensor theories from $\Lambda(\phi)$ Plebanski gravity}

\author{David Beke\footnote{david.beke@ugent.be}}
\affiliation{Department of Mathematical Analysis EA16, Ghent University, 
Galglaan 2, 9000 Gent, Belgium}

\date{\today}
\begin{abstract}
We study a modification of the Plebanski action, which generically corresponds to a bi-metric theory of gravity, and identify a subclass which is equivalent to the Bergmann-Wagoner-Nordtvedt class of scalar-tensor theories. In this manner, scalar-tensor theories are displayed as constrained BF theories. We find that in this subclass, there is no need to impose reality of the Urbantke metrics, as also the theory with real bivectors is a scalar-tensor theory with a real Lorentzian metric. Furthermore, while under the former reality conditions instabilities can arise from a wrong sign of the scalar mode kinetic term, we show that such problems do not appear if the bivectors are required to be real. Finally, we discuss how matter can be coupled to these theories. The phenomenology of scalar field dark matter arises naturally within this framework. 
\end{abstract}
\maketitle

\section{Introduction}

This paper is concerned with scalar-tensor (ST) theories, i.e. theories with the following action principle \cite{Bergmann, Wagoner, Nordtvedt}:
\be
S[g_{\m\n},\psi,Q]=\int \psi R(g_{\m\n}) +{\mathcal L}_{\psi}(g_{\m\n},\psi)+{\mathcal L}_m(\psi,g_{\m\n},Q). \label{ST}
\ee
The dynamical variables are the metric tensor field $g_{\m\n}$, the scalar field $\psi$ and tensor matter fields $Q$.\footnote{We will restrict the discussion to tensor matter fields, but since we will be considering a first order formalism for gravity, the connection being a dynamical variable, spinor matter fields are easily described.}
Historically, such theories have attracted interest because of the possibility of a time-varying gravitational constant $\psi$. The most natural theory to consider from this point of view is the Jordan-Brans-Dicke (JBD) choice 
\be
{\mathcal L}^{\text (JBD)}_{\psi}=-\frac{\omega}{\psi}\psi_{,\m}\psi_{,\n}g^{\m\n}, \quad {\mathcal L}^{\text (JBD)}_m(\psi,g_{\m\n},Q) ={\mathcal L}_m(g_{\m\n},Q),
\ee
with a dimensionless coupling constant $\omega$. This theory is severely constrained by solar-system experiments, the most up-to date constraint being $\omega>4\times 10^4$ \cite{Perivolaropoulos}. Recently there has nevertheless been a renewed interest in ST theories \cite{Fujii}, incited both by theoretical and phenomenological results. On the theoretical front, the dilaton arises from dimensional reduction, as has been known since the study of Kaluza-Klein theory. Most notably the low energy limit of string theory is a ST theory \cite{Green}. On the phenomenological front, the interest has been spurred by the possibility of a first order phase transition ending an inflationary epoch, providing a graceful exit and avoiding the slow-roll conditions \cite{Faraoni}. One is however forced to consider more general scalar actions $\mathcal{L}_\psi$, by introducing a potential $V(\psi)$ and by allowing $\omega$ to depend on the scalar field $\omega(\psi)$. Such a theory has the potential to combine ST phenomenology at early times with present constraints, whenever $\omega$ evolves from small values to the large-value general relativity limit. Such a mechanism has also been displayed in a class of ST theories in which all dilaton coupling functions coincide \cite{Damour1994}, cosmological evolution attracting the dilaton to a state in which it decouples. 

It will be shown in this paper how massive ST theories arise in $\Lambda(\phi)$ Plebanski theory. This class of modified gravity theories is the non-chiral analogue of non-metric gravity \cite{Krasnov2008, Bengtsson}, which has been introduced motivated by the asymptotic safety program. Aside from providing an arena to examine effective quantum gravity actions, these theories offer a promising avenue for a unified description of gravity and Yang-Mills theories \cite{Smolin2009,Lisi2010,Torrez,KrasnovSymmBr}. Both modified theories are inspired by the Plebanski formalism for general relativity, which describes gravity by a cubic action principle and more in particular as a constrained $SU(2)$ or $SO(3,1)$ BF-theory. The fundamental geometric variable in this formalism is a $\mathfrak{su}(2)$ respectively a $\mathfrak{so}(3,1)$ bivector, and the Lorentzian metric is obtained as a derived geometric quantity. While the chiral and non-chiral Plebanski formalism are equivalent, both being a formalism for regular general relativity, the non-chiral modified theory differs significantly from non-metric gravity. Generically $\Lambda(\phi)$ Plebanski theory contains 8 propagating degrees of freedom (DOFs), as opposed to the 2 DOFs of non-metric gravity \cite{Alexandrov2009,Speziale2010}. We will identify a non-generic subclass in which these 8 DOFs reduce to the 3 DOFs of a ST theory. The scalar action $\mathcal{L}_\psi$ that arises naturally from these theories behaves as a free field for small values and grows exponentially for large values. Minimally coupled scalar fields with such a potential have attracted interest because of the dark matter phenomenology they produce \cite{Matos2000,Sahni2000}. 

Furthermore, in this ST subclass it is possible to discuss the fundamental aspects of $\Lambda(\phi)$ Plebanski theory in detail. In particular, different reality conditions can be imposed on the bivectors such as to obtain a real Lorentzian metric, and in the unmodified theory these are equivalent. In the modified theory, different reality conditions yield a different theory, and it is not clear which reality conditions to impose. We will compare these different conditions in the ST subclass. Another aspect of $\Lambda(\phi)$ Plebanski theory which has not been clarified is the coupling of matter fields. Without an understanding of this coupling, it is impossible to extract physical consequences already in the ST subclass: in the absence of matter fields, it is possible to make a conformal field redefinition $\hat{g}_{\m\n}=\psi g_{\m\n}$, and have a minimally coupled scalar field. It is the presence of non-conformally invariant matter fields (such as a massive scalar point particle) which singles out the physical metric as the one with respect to which test particles move on geodesics. We discuss a method to couple matter to $\Lambda(\phi)$ Plebanski theory by hand, and show how minimal coupling is obtained for the ST theories. Whether minimal coupling arises as an attractor in phase space is left as an open question. This `coupling by hand' is to be contrasted with the more ambitious approach followed in \cite{KrasnovSymmBr}, where it is attempted to unify not only Yang-Mills fields but also fermionic matter fields with the gravitational fields. The matter coupling discussed in this paper is more pragmatic, but the results are not without interest for this unified perspective. We are able to single out the metric with respect to which the scalar field is minimally coupled, establishing how at least one of the emerging matter fields couples to gravity.

After a review of $\Lambda(\phi)$ Plebanski theory in section \ref{sec:PlebSec}, the subclass of the theory resulting in ST theory will be discussed in section \ref{sec:ST}. In section \ref{sec:MC} we discuss the coupling to matter. Finally, aside from the physical motivations discussed above, the results of this paper can also be interpreted as giving a BF formulation of massive ST theories. Massless ST theories cannot be obtained in the $\Lambda(\phi)$ Plebanski framework, but finding a BF formulation of massless ST theory is straightforward once the description of massive ST theories is understood. We discuss this in section \ref{sec:JBD}, and show how the JBD theory, which doesn't arise naturally in this setting, can be recovered. 


\section{$\Lambda(\phi)$ Plebanski theory}\label{sec:PlebSec}

In the non-chiral Plebanski formalism, Lorentzian, resp. Euclidean, general relativity is described by a set of $\mathfrak{so}(3,1)$, resp. $\mathfrak{so}(4)$-valued bivectors $B^{IJ}_{\mu\nu}$ (i.e. $B^{IJ}_{\mu\nu}=-B^{JI}_{\mu\nu}=-B^{IJ}_{\nu\mu}$). We refer to appendix \ref{App:Not} for further details on the notation. The Plebanski action principle
\be
S_{Pl}\left[B^{IJ}_{\mu\nu},A_\mu^{IJ},\phi_{IJKL}\right] = \int B^{IJ}\wedge F_{IJ}(A)-\frac{1}{2}\left(\phi_{IJKL}+\f\Lambda3\eps_{IJKL}\right)B^{IJ}\wedge B^{KL},\label{Pleb}
\ee
is that of a constrained BF theory. Here, \`{a} la Palatini, the connection $A_\mu^{IJ}$ is taken as an independent variable, and $\phi_{IJKL}$ is a Lagrange multiplier obeying the symmetries of the Riemann tensor, $\phi_{IJKL}=\phi_{KLIJ}=-\phi_{JIKL}$ and $\eps^{IJKL}\phi_{IJKL}=0$. The role of the simplicity constraints
\be
B^{IJ}\wedge B^{KL} - \sigma\frac{1}{24}\eps^{IJKL}\eps_{MNPQ}B^{MN}\wedge B^{PQ} = 0,\label{SimplCons}
\ee
which result from varying with respect to the Lagrange multipliers, is to put these bivectors into correspondence with a tetrad. This correspondence occurs because the constraints are solved, as will be detailed below, by
\be
B_{\text{gr}}^{IJ}=\pm \left(\star e\wedge e\right)^{IJ}, \quad B_{\text{top}}^{IJ} = \pm e^I\wedge e^J,\label{Bsol}
\ee
where $\star$ is the hodge dual in the Lie algebra (not the space-time hodge dual). The remaining field equations 
\bea
d_A B & = & 0\label{Bcompat}\\
F^{IJ}[A] & = & \phi^{IJ}{}_{KL}B^{KL},\label{PlebEinst}
\eea
with $d_A$ the connection differential, can be used to show that (i) $A$ is the spin connection $e^I_\nu\nabla_\mu e^{\nu J}$, with $\nabla$ the metric-compatible derivative, (ii) for the solution $B_{\text {gr}}$, the metric satisfies Einstein's equations, and (iii) for the solution $B_{\text {top}}$, a topological theory is obtained \cite{Smolin2009}. These results are also apparent from substituting \Ref{Bsol} back in the action, yielding the Einstein-Cartan action for $B=B_{\text {gr}}$ and the topological action $\int e^I\wedge e^J\wedge F[A]_{IJ}$ for $B=B_{\text{top}}$.

The correspondence between BF theory and general relativity that is provided by the Plebanski formalism rests on the simplicity constraints. These constraints break a part of the gauge symmetry of BF theory, and without them the theory is topological. They can however also be relaxed by replacing $\phi_{IJKL}$ by $\phi_{IJKL}+\f16\Lambda(\phi)\eps_{IJKL}$, where $\Lambda$ is a function of scalar invariants of the Lagrange multiplier.\footnote{Studying such relaxations is motivated by the effective action rationale of allowing all terms of correct dimension and symmetry. We refer to \cite{Krasnov2010,Krasnov2012} for a more complete discussion of the motivations.} The gauge symmetry of BF theory is still partially broken, and the resulting theory, with action principle 
\be
S _{MP}=  \int B^{IJ}\wedge F_{IJ}- \frac{1}{2}\left(\phi_{IJKL}+\frac{\Lambda(\phi)}{6}\eps_{IJKL}\right)B^{IJ}\wedge B^{KL},\label{ModifPleb}
\ee
is not topological. The field equations (\ref{SimplCons},\ref{Bcompat},\ref{PlebEinst}) are then modified to
\bea
 B^{IJ}\wedge B^{KL}-\f1{12}\eps^{IJKL} <B\wedge\star B> & = & -\f13\f{\partial\Lambda}{\partial\phi^{IJKL}}<B\wedge\star B>,\label{ModifSimp}\\
 d_AB & = & 0,\label{Bcompat2}\\
 F^{IJ}(A) & = & \left(\phi^{IJ}{}_{KL} + \f{\Lambda(\phi)}6 \eps^{IJ}{}_{KL}\right)B^{KL}. \label{ModEinst}
 \eea
When the Hessian of $\Lambda(\phi)$ is non-singular, \Ref{ModifSimp} yields equations for the Lagrange multipliers instead of genuine constraints. It was shown in \cite{Alexandrov2009} that, for such non-singular cases, the resulting theory contains 8 DOFs. In \cite{Speziale2010,Beke2012}, those DOFs were identified as a massless graviton coupled to a massive graviton accompanied by the Boulware-Deser ghost. Apart from the 2 DOF general relativity case \Ref{Pleb}, $\Lambda(\phi)=\Lambda$, singular cases yielding less than 8 DOFs non-perturbatively have not been discussed in the literature. In the next section we will show how a singular potential, where $\Lambda(\phi)$ is only a functional of $\varphi_0=\tr(\phi)$, gives rise to a theory with 3 DOFs: a massless graviton coupled to a massive scalar field. 

Before discussing this, it is useful to explain in more detail how to solve the simplicity constraints. In order to do so, we decompose \Ref{SimplCons} in its irreducible components. 
First, we split $B^{IJ}$ in its Lie-algebra selfdual (SD) and anti-selfdual (ASD) components: $B^{IJ}=P^{IJ}_{(+)i}B^{i}_{(+)} + P^{IJ}_{(-)i}B^i_{(-)}$. In the Euclidean case, one has introduced 6 real bivectors $B^i_{(+)}$, $B^i_{(+)}$, while in the Lorentzian case 3 complex bivectors $B^{i}_{(+)}$ are introduced, and in order for $B^{IJ}$ to be real, the ASD part is its complex conjugate, 
\be
B^i_{(-)}=B^{i\;*}_{(+)}.\label{Breal}
\ee
Any set of 3 bivectors can be required, as a condition on the space-time metric, to be (A)SD wrt the space-time hodge dual\footnote{Whenever $B^i\wedge B^j$ is non-degenerate this condition determines the metric uniquely up to conformal factor.}, such that $B$ can be parametrized by \cite{Freidel2008,Speziale2010}
\be
B^{IJ}=P^{IJ}_{(+),i}b^{(+)}{}^i_j\Sigma^j_{(+)\mu\nu}[e^{(+)}{}^I_{\mu}]+P^{IJ}_{(-),i}b^{(-)}{}^i_j\Sigma^j_{(-)\mu\nu}[e^{(-)}{}^I_{\mu}].\label{BparTemp}
\ee
Here $B^i_{(-)}$ has been required to be ASD instead of SD, such as to be consistent with \Ref{Breal} and a real hodge dual, in the Lorentzian case. In terms of these parameters, the Lorentzian reality conditions are given by
\be
b^{(-)} = b^{(+)\; *} , \quad e^{(-)}={e^{(+)\;*}}, \label{realities}
\ee
while for the Euclidean case all fields are taken to be real. One should note that the parametrization \Ref{BparTemp} is not unique: $B^{IJ}$ is invariant under a transformation $b\rightarrow \Omega^2 b$, $e^I_\mu\rightarrow \Omega^{-1}e^I_\mu$. In order to fix this ambiguity, one requires $\det(b)=1$, and in the Euclidean case introduces $\eta,\bar{\eta}=\pm 1$ to take into account that this transformation cannot change the sign of $\det(b)$:
\be
B^{IJ}=\tilde{\eta}P^{IJ}_{(+),i}b^{(+)}{}^i_j\Sigma^j_{(+)\mu\nu}[e^{(+)}{}^I_{\mu}]+ \tilde{\eta}\eta P^{IJ}_{(-),i}b^{(-)}{}^i_j\Sigma^j_{(-)\mu\nu}[e^{(-)}{}^I_{\mu}].\label{BDecomp}
\ee
In the Lorentzian case, $\eta=\tilde{\eta}=1$. Using this parametrization, the irreducible decomposition of \Ref{SimplCons} is given by
\bea
&&{\bf \Pi}^{IJKL}_{{\bf(2,0)},ij}2\s\sqrt{\s}e^{(+)}m^{(+)ij} - {\bf \Pi}^{IJKL}_{{\bf(0,2)},ij}2\s\sqrt{\s}e^{(-)}{}m^{(-)ij} \label{SimplConsIrr}\\
&& \quad\quad + {\bf \Pi}^{IJKL}_{{\bf(1,1)},ij}4\eta b^{(+)}{}^i_kb^{(-)}{}^j_l\Sigma^k_{(+)}\wedge\Sigma^l_{(-)} + {\bf \Pi}^{IJKL}_{{\bf(0,0)}} 2\sqrt{\s}(e^{(+)}m^{(+)}-e^{(-)}m^{(-)}) = 0, \nonumber
\eea
where $m^{(+)ij}=b^{(+)}{}^i_kb^{(+)}{}^j_k$, $m^{(+)}=\tr(m^{(+)ij})$, $e^{(+)}=\det(e^{(+)}{}^I_\m)$ and likewise for the ASD quantities.
It follows, using unimodularity, from the $\bf (2,0)$- and the $\bf (0,2)$-components that\footnote{More precisely, in the Lorentzian case also $m^{(+)}_{ij}=e^{\pm i\f{2\pi}3}\d_{ij}$ are unimodular solutions. However the condition $\det{b}=1$ only fixes $\Omega^2$ up to a factor $e^{\pm i\f{2\pi}3}$, and this remaining ambiguity can be used to fix $m^{(+)}_{ij}=\d_{ij}$.} 
\be
m^{(+)ij}= m^{(-)ij} = \d^{ij},
\ee
such that the scalar component reduces to 
\be
e^{(+)}=e^{(-)}.\label{ConfEqual}
\ee
The $({\mathbf 1},{\mathbf 1})$-components are solved by $e^{(-)I}_{\mu}\propto e^{(+)I}_{\mu}$, up to a proper Lorentz transformation which doesn't affect the result. In the Euclidean case, it follows from \Ref{ConfEqual} that this factor is $\pm1$:
\be\label{eSolso4}
\mathfrak{so}(4):    \quad e^{(-)}{}^I_{\mu} = \pm e^{(+)}{}^I_{\mu} \equiv e^I_\m.
\ee
In the Lorentzian case, the factor is any 4th root of 1,
\be
\mathfrak{so}(3,1): \quad e^{(-)}{}^I_{\mu} = e^{ik\f\pi2} e^{(+)}{}^I_{\mu}, \quad k =0,...,3. 
\ee
The reality conditions \Ref{realities} then imply that $e^{(+)}{}^I_\m$ and $e^{(-)}{}^I_\m$ can be related to a real tetrad $e^I_\m$ by
\be\label{eSolso31}
e^{(\eps)}{}^I_\m = \pm e^{i\eps n\f\pi4} e^I_\m, \quad n=0,...,3.
\ee

Substituting $b^i_j=\bar{b}^i_j=\delta^i_j$ and \Ref{eSolso4} resp. \Ref{eSolso31} in the decomposition (\ref{BDecomp}), one obtains for the Euclidean and Lorentzian case that\be
B^{IJ}=\tilde{\eta}\frac{1+\eta}{2}e^I\wedge e^J +\tilde{\eta}\frac{1-\eta}{2}   (\star e\wedge e)^{IJ},\label{Bmetr}
\ee
where for the Lorentzian case we have conveniently replaced $n=0,...,3$ by $\eta=\pm1$, $\tilde{\eta}=\pm 1$ to parametrize the different solutions, such as to explicitly show the equivalence of the Euclidean and the Lorentzian case. The explicit relation between these parameters is given by $n=\left(\tilde{\eta}+\f{\eta+1}{2}\right)Ê{\text{mod} \;4}$.
For $\eta=1$, one obtains the topological solution of (\ref{Bsol}), while for $\eta=-1$ the gravitational solution is found.

Let us point out that Lorentzian gravity, with a real Lorentzian metric, can be recovered in two different ways. In the first approach, the bivectors $B$ are taken to be real, as discussed above. Alternatively, one can require the reality of $e^{(+)}$ and $e^{(-)}$, after introducing $\eta$ and $\tilde{\eta}$ in the parametrization, i.e. impose the reality conditions of the selfdual Plebanski formulation. Since under this assumption both real tetrads are independent, the results can be obtained by studying the Euclidean case and introducing factors $i$ where appropriate. If the simplicity constraints are satisfied, this reality condition gives either real (topological sector) or purely imaginary bivectors (gravitational sector). Factoring out $i$ in the action of the latter sector, it is found that the two reality conditions are equivalent, {\emph{in the presence of the simplicity constraints}}. In the absence of the simplicity constraints, the phase of $B$ is dynamical, and cannot simply be factored out of the action. Hence these reality conditions differ, and yield a different $\Lambda(\phi)$ Plebanski theory. We will clarify the differences between both reality condition in the ST subclass that is discussed in the next section. Under the condition of real $e^{(+)}$ and $e^{(-)}$, the theory splits in two sectors, one having the right sign and one having the wrong sign of the kinetic term. Under the condition of real $B$, the scalar field is guaranteed to have the right sign of the kinetic term in both sectors.


\section{Massive Scalar Tensor theory from modified constraints}\label{sec:ST}

A `weak' relaxation of the simplicity constraints, is to relax only the scalar component in (\ref{SimplConsIrr}), i.e. relax (\ref{ConfEqual}). In that case $e^{(+)}{}^I_{\mu}$ and $e^{(-)}{}^I_{\mu}$ are constrained to be conformally related:
\be
m^{ij}=\bar{m}^{ij}=\delta^{ij}, \quad e^{(-)}{}^I_{\mu}=\psi e^{(+)}{}^I_{\mu}. \label{ConstrSols}
\ee
In the Euclidean case, $\psi$ is simply a real variable, while in the Lorentzian case, it follows from \Ref{realities} that it is a phase $\psi=e^{i\phi}$ and one can relate $e^{(+)}$ and $e^{(-)}$ to a real tetrad $e$ by $e^{(+)}{}^I_\mu=\pm e^{-i\f\phi2}e^I_\m$, $e^{(-)}{}^I_\mu=\pm e^{i\f\phi2}e^I_\m$. $B$ is given by
\bea
\mathfrak{so}(4) :   &&     B^{IJ}=\tilde{\eta}\frac{1-\eta \psi^2}{2} \left(\star e\wedge e\right)^{IJ} + \tilde{\eta}\frac{1+ \eta \psi^2}{2} e^I\wedge e^J; \label{BSO4}\\
\mathfrak{so}(3,1): &&     B^{IJ}= - \sin(\phi) \left(\star e\wedge e\right)^{IJ} - \cos(\phi) e^I\wedge e^J. \label{BSO31}
\eea
This solution is equivalent, after a redefinition of the tetrad $\sqrt{\left|\tilde{\eta}\frac{1-\eta \psi^2}{2}\right|}e^I_\m\rightarrow e^I_\m$ resp. $\sqrt{\left|\sin(\phi))\right|}e^I_\m\rightarrow \tilde{e}^I_\m$, and of the scalar $\frac{1+ \eta \psi^2}{1-\eta\psi^2}\rightarrow \tilde{\psi}$, resp. $\cot(\phi)\rightarrow \tilde{\psi}$, to
\be
B^{IJ}=\pm\left(\star+\tilde{\psi}\right)\tilde{e}^I\wedge \tilde{e}^J,\label{BSTgen}
\ee
which was also discussed in \cite{Capovilla2001,Smolin2010}, where an extra constraint was imposed to fix $\tilde{\psi}$ to a constant and obtain the Holst action, $\tilde{\psi}$ playing the role of the (inverse) Immirzi parameter. We will show that if this constraint is not imposed, $\tilde{\psi}$ becomes a scalar propagating degree of freedom, responsible for the ST character of the theory. In this sense, the ST theories obtained here originate from considering a time-varying Immirzi parameter, rather than a time-varying gravitational constant. For the generic analysis of section \ref{STsubs} we will however stick to the parametrization (\ref{BSO4}-\ref{BSO31}): in this parametrization, one can use \Ref{ConstrSols} to split equation \Ref{Bcompat2} in its (A)SD parts most elegantly. 

In the class of theories described by (\ref{ModifPleb}), this relaxation can be achieved by choosing
\be
\Lambda(\phi)=\Lambda_0(\varphi_0),\label{ScalarPotential}
\ee
as the right-hand side of equation \Ref{ModifSimp} then only contributes to the $\bf (0,0)$-component, leaving the  $\bf (2,0)$-, $\bf (0,2)$- and $\bf (1,1)$-components of the constraints untouched. Hence these are solved by \Ref{ConstrSols}. In comparison to the theory with a generic potential $\Lambda(\phi)$, where also these components are relaxed, this subclass is more constrained. More specifically, as this generic case corresponds to bi-metric theories of gravity \cite{Speziale2010,Beke2012}, the ST subclass can be interpreted as a constrained bi-metric theory, where the 2 metrics are constrained to be conformally related.

The scalar component on the other hand, is modified to
\be
\Lambda'_0(\varphi_0) =-\f12 \f{<B\wedge B>}{<B\wedge\star B>} = -\f{\s\tilde{\psi}}{1+\s\tilde{\psi}^2}.\label{TrEq}
\ee
Generically, $\Lambda''_0(\varphi_0)\neq 0$ and this equation can be solved for $\varphi_0$. Values of $\varphi_0$ at which $\Lambda''(\varphi_0)=0$ then serve as singular points of phase space, splitting the theory in different sectors. Before discussing this, we first consider the case $\Lambda_0''(\varphi_0)\equiv 0$, the scalar component still constraining the dynamical variables.

\subsection{Linear potential $\Lambda_0$}\label{sec:Lin}
If the scalar potential $\Lambda_0$ is given by
\be
\Lambda_0(\varphi_0)=2\Lambda_0-\f1{4a}\varphi_0,\label{LinPot}
\ee
equation \Ref{TrEq} can be solved for $\tilde{\psi}$ iff $|a|\geq  \f\s2$ (otherwise the action has no stationary points). The solution is given by
\be
\tilde{\psi} (a)= 2a \pm \sqrt{4a^2-\s}. \label{PsiSol}
\ee
Substitution of \Ref{BSTgen} and \Ref{PsiSol} in the action yields 
\be
S=\int\pm\star\tilde{e}^I\wedge\tilde{e}^J\wedge F_{IJ} +\tilde{\psi}(a)\tilde{e}^I\wedge\tilde{e}^J\wedge F_{IJ}- 4\Lambda_0 \left(1+\s\tilde{\psi}^2(a)\right)\tilde{e}, \label{HolstCosm}
\ee
which is the Holst action, with the Immirzi parameter and the cosmological constant given by
\bea
\gamma^{-1} & = &  \tilde{\psi}(a) = 2a \pm \sqrt{4a^2-\s},\label{Imm}\\
\Lambda & = & \Lambda_0\left(1+\s\tilde{\psi}^2(a)\right) = 4a\Lambda_0\sigma\left(2a \pm \sqrt{4a^2-\s}\right).\label{Cosm}
\eea
In \Ref{HolstCosm}, the $\pm$ signs correspond to 2 different sectors of the theory. In the absence of matter, the effect of the sign is equivalent to switching the sign of the potential term (which can be accounted for by changing the sign of the parameters appearing in the potential), and in the remainder of this section we discuss without loss of generality the + sector. In the presence of matter fields, the signs of all kinetic terms have to be equal, and in section \ref{sec:MC} both sectors will be considered.

It has been noted in \cite{Capovilla2001,Smolin2010}, that \Ref{Pleb} with $\phi$ satisfying 
\be
(\delta_{IJKL}+a\s\eps_{IJKL})\phi^{IJKL}=0, \quad |a|>\f\s2,\label{SkewCond}
\ee
instead of $\eps_{IJKL}\phi^{IJKL}=0$, corresponds precisely to (\ref{HolstCosm}-\ref{Cosm}) (without the $\pm$ signs). This point of view is easily understood to be equivalent to \Ref{ModifPleb}, with the potential $\Lambda$ given by \Ref{LinPot} and the usual condition $\eps_{IJKL}\phi^{IJKL}=0$. Starting with \Ref{SkewCond}, one can make a field redefinition $\phi'=\phi+\f1{24a}\varphi_0\eps$ satisfying $\eps\cdot\phi' = 0$. In terms of this variable, the action \Ref{Pleb} is then given by
\be
\int B^{IJ}\wedge F_{IJ} - \f12\left(\phi'_{IJKL}+\f16(2\Lambda-\f1{4a}\varphi'_0)\eps_{IJKL}\right)B^{IJ}\wedge B^{KL}.
\ee

\subsection{Generic potential $\Lambda_0$}\label{STsubs}

\subsubsection{The Euclidean case}

As already mentioned, the compatibility equation is most elegantly solved by parametrizing the solution of the unmodified simplicity components by \Ref{BSO4}. In terms of these parameters, \Ref{TrEq} reads
\be
\Lambda'_0(\varphi_0) =-\f12\frac{1-\psi^4}{1+\psi^4}.
\ee
In the non-linear case, the equation can be solved for $\varphi_0(\psi)$. In general, this solution is not unique, different solutions yielding $n$ different sectors in which $\varphi_0\in I_i=\left]\varphi_i,\varphi_{i+1}\right[$, $i=1..n$ with $\varphi_1=-\infty$, $\Lambda''(\varphi_j)=0$, $j=2..n$ and $\varphi_{n+1}=\infty$, and thus $-\f12\f{1-\psi^4}{1+\psi^4}\in \Lambda'_0(I_i)$. To avoid heavy notation, we will study the theory in one such sector, i.e. choose one (arbitrary) solution. The simplest case is the quadratic potential
\be
\Lambda_0(\varphi_0)=\frac{a}{16}\varphi_0^2,\label{LambdaQuad}
\ee
for which a unique solution
\be
\varphi_0=-\frac{4}{a}\frac{1-\psi^4}{1+\psi^4}\label{TraceInQuad}
\ee
exists. We will use this case for illustrative purposes. 

Substitution of equation (\ref{BSO4}) in equation \Ref{Bcompat2}, and splitting the equation and the variables in their SD and ASD parts yields
\bea
d\Sigma_{(+)}^i-\eps^i_{jk}A^j_{(+)}\wedge \Sigma^k_{(+)} & = & 0\\
d\left(\psi^2\Sigma_{(-)}^i\right)-\psi^2\eps^i_{jk}A^j_{(-)}\wedge \Sigma^k_{(-)} & = & 0
\eea
These equations can be solved for $A_{(\eps)\mu}^{i}$, yielding
\bea
A^j_{(+)\mu} & = & P^j_{(+)IJ}e^I_{\nu}\nabla_{\mu} e^{\nu J}\\
A^j_{(-)\mu} & = & P^j_{(-)IJ}\psi e^I_{\nu}\bar{\nabla}_{\mu}\left(\frac{1}{\psi} e^{\nu J}\right)\\
                       & = & P^j_{(-)IJ}\left(e^I_{\nu}\nabla_{\mu} e^{\nu J}+\frac{2}{\psi}e_{\mu}^Ie^{\nu J}\nabla_{\nu}\psi\right)
\eea
where $\nabla$ (resp. $\bar{\nabla}$) is the covariant derivative wrt the metric associated to $e^I_{\mu}$ (resp. $\psi e^{I}_{\mu}$). Hence we have solved for the connection:
\be
A_{\mu}^{IJ}=e^I_{\nu}\nabla_{\mu}e^{\nu J} + \frac{2}{\psi}P^{IJ}_{(-)\;KL}e^K_{\mu}e^{L}_{\nu}\nabla^{\nu}\psi. \label{ConnScalTetr}
\ee

Substituting (\ref{BSO4}), the solution of (\ref{TrEq}), and (\ref{ConnScalTetr}) in the action, one obtains, up to a total derivative,
\be
\int e\left[\tilde{\eta}\frac{1-\eta \psi^2}{2}R-3\tilde{\eta}\eta\nabla_{\kappa}\psi\nabla^{\kappa}\psi- V_0(\psi)\right],\label{EffAct}
\ee
where 
\be
V_0(\psi)= \Lambda_0(\varphi_0(\psi))(1+\psi^4)+\f 1 2 \varphi_0(\psi)(1-\psi^4).
\ee
As can be expected, the choice of (non-linear) $\Lambda_0$ only affects the potential terms, the structure of phase space is not affected by it. For the choice (\ref{LambdaQuad}),  
\be
V_0(\psi)=-\frac{1}{a}\frac{1-\psi^4}{1+\psi^4}\left(1-\psi^4\right).
\ee
A more direct way of deriving \Ref{EffAct} is by substituting \Ref{ConstrSols} in the action (67) of \cite{Freidel2008}, where the compatibility equation \Ref{Bcompat2} was solved for general $B$, using the parametrization \Ref{BDecomp}, before using \Ref{ConstrSols}. 

One can diagonalize the kinetic terms by switching from the Jordan frame to the Einstein frame, i.e. making a conformal field redefinition $\hat{g}_{\mu\nu}=\Omega^{-2}g_{\mu\nu}$, with 
\be
\Omega = \sqrt{\frac{2\tilde{\eta}}{1-\eta \psi^2}},\label{ConfRedef}
\ee
where again we can, without loss of generality, choose $\tilde{\eta}$ such that the argument of the square root is positive. 
The action is then given, up to total derivatives, by (see e.g. \cite{Fujii})
\be
\int e \left[R - \frac{6\eta}{(1-\eta\psi^2)^2}\nabla_{\mu}\psi\nabla^{\mu}\psi-\frac{4}{(1-\eta\psi^2)^2}V_0(\psi)\right].
\ee
The canonical kinetic term for the scalar field is obtained by making a field redefinition $\hat{\psi}(\psi)$ with $\f{d\hat{\psi}}{d\psi}=\f{\sqrt{6}}{1-\eta\psi^2}$, i.e.
\be
\hat{\psi}(\psi) = 
\left\{
\begin{array}{lr}
\sqrt{6}\arctan(\psi),             & \eta=-1\\
\sqrt{6}\arctanh(\psi),           & \eta=1,|\psi|<1 \\
\sqrt{6}\arctanh(\psi^{-1})  & \eta=1,|\psi|>1
\end{array}\right..
\ee
It is useful to note that in the first case, the scalar field will be constrained to $\hat{\psi} \in ]-\sqrt{6}\frac{\pi}{2},\sqrt{6}\frac{\pi}{2}[$. Explicitly, the action in the respective sectors is given by
\be\label{SFinal}
S=\left\{
\begin{array}{lr}
 \int e \left[ R + \nabla_{\mu}\psi\nabla^{\mu}\psi - V_{\rm gr}(\psi)\right],             & \eta=-1\\
 \int e \left[ R - \nabla_{\mu}\psi\nabla^{\mu}\psi - V_{\rm{top}+}(\psi)\right],            & \eta=1,|\psi|<1 \\
 \int e \left[ R - \nabla_{\mu}\psi\nabla^{\mu}\psi - V_{\rm{top}-}(\psi)\right],            & \eta=1,|\psi|>1
\end{array}\right..
\ee
with
\bea
V_{\rm gr}(\psi)       & = & 4\cos^4(\frac{\psi}{\sqrt{6}})V_0\left(\tan(\f\psi{\sqrt{6}})\right)\\
V_{\rm{top}+}(\psi) & = & 4\cosh^4(\frac{\psi}{\sqrt{6}})V_0\left(\tanh(\frac{\psi}{\sqrt{6}})\right)\label{Vtop+}\\
V_{\rm{top}-}(\psi)  & = & 4\sinh^4(\frac{\psi}{\sqrt{6}})V_0\left(\tanh^{-1}(\frac{\psi}{\sqrt{6}})\right)\label{Vtop-}
\eea
Note that all sectors have three propagating degrees of freedom. This is also true for the case $\eta=1$, which corresponds to a topological theory when all constraints are kept.
The sector with $\eta=-1$ has the wrong sign of the kinetic term. In the sectors with $\eta=1$, The kinetic terms of the graviton and the scalar field have the same sign, and the theory is ghost-free in the absence of matter fields.

Consider the potential \Ref{LambdaQuad}, with $\eta=1$, as an illustration. The potentials (\ref{Vtop+},\ref{Vtop-}) reduce to
\be
V_{\mathrm{quad}}(\psi)=-\frac{4}{a}\frac{\left(\sinh^4(\frac{\psi}{\sqrt{6}})-\cosh^4(\frac{\psi}{\sqrt{6}})\right)^2}{\left(\sinh^4(\frac{\psi}{\sqrt{6}})+\cosh^4(\frac{\psi}{\sqrt{6}})\right)}.\label{PotQuad}
\ee
This potential is bounded, but we will see how the Immirzi parameter and the cosmological constant affect this result in section \ref{sec:Imm}.

\subsubsection{The Lorentzian case}

The Lorentzian analysis is analogous to the Euclidean one, the difference lying in the details. Equation \Ref{TrEq} now reduces to
\be
\Lambda_0'(\varphi_0) = - \f 12 \tan(2\phi),\label{ModifScalLor}
\ee
which generically can be solved for $\varphi_0(\phi)$, again with the possibility of different solutions from which we pick one. For the simplest choice \Ref{LambdaQuad}, the solution is given by $\varphi_0=-\f4a\tan(2\phi)$. The (A)SD parts of the compatibility equations can again be solved for the (A)SD parts of the connection, the solutions being given by
\be
A^{(\eps)}{}^I_\m = - P^{(\eps)i}_{IJ}\left[e^I_\n\nabla_\m e^{\n J}-\eps i e^I_\m e^{\n J}\nabla_\n\phi\right], 
\ee
such that
\be
A_\m^{IJ} = e^I_\n\nabla_\m e^{\n J} +\f12 \eps^{IJ}{}_{KL}e^K_\m e^{\n L}\nabla_\n\phi.\label{ConnLor}
\ee
The resulting ST action is given by
\be
\int e\left[ -\sin(\phi) R + \f32 \sin(\phi)\nabla_\m\phi\nabla^\m\phi - V_0(\phi)\right],\label{LorSTJordan}
\ee
with
\be
V_0(\phi) = -\varphi_0(\phi)\sin(2\phi) - 2\Lambda_0(\varphi_0(\phi))\cos(2\phi),
\ee
the choice \Ref{LambdaQuad} leading to
\be
V_0(\phi) = \f2a\f{\sin^2(2\phi)}{\cos(2\phi)}.\label{LorV0}
\ee
The sign of the scalar and tensor kinetic terms is always equal, which is contrary to the situation in the Euclidean theory. This is clearly not changed by going to the Einstein frame $\hat{g}_{\m\n} = |\sin(\phi)|g_{\m\n}$, in which the action is given by (immediately dropping the hats, and again dropping the $\pm$ signs by considering $-\sin(\phi)>0$)
\be
S=\int e\left[ R -\f 3 2 \f1{\sin^2(\phi)}\nabla_{\m}\phi\nabla^{\m}\phi - \f1{\sin^2(\phi)}V_0(\phi)\right].
\ee
When $V_0$ is bounded from below, the Hamiltonian also is bounded from below, in the absence of matter fields. 

The canonical scalar kinetic energy term 
\be
S=\int e\left[R-\nabla_\m\psi\nabla^\m\psi-V(\psi)\right], \quad V(\psi) = \f 1 {\sin^2(\phi(\psi))}V_0(\phi(\psi))\label{SFinalLor}
\ee
can be obtained by taking $\psi=\sqrt{\f32}\ln(|\tan(\f\phi2)|)$. Again the original theory contains two sectors: this field transformation maps both $]-\pi,0[$ and $]0,\pi[$ to $\mathbb{R}$. In general they differ by the resulting potential, but with \Ref{LorV0} being a symmetric function in $\tan(\f\phi2)$, for \Ref{LambdaQuad} both theories are equivalent, and the potential is given by
\be
V(\phi)=\f8a\f{\sinh^2(\sqrt{\f23}\psi)}{\sinh^2(\sqrt{\f23}\psi)-1}.\label{LorV}
\ee 
This potential approaches $\f8a$ for large $\psi$, and has a singularity at $\psi=\pm\sqrt{\f32}\ln(1+\sqrt{2})\equiv \pm c$, effectively splitting the theory in 3 sectors (fields taking value in one of the three regions split by these singularities, stay in this region). There only is a minimum (resp. maximum) in the bounded sector $\psi\in]-c,c[$, for negative (resp. positive) $a$. In the next section we will find that this situation is changed when the Immirzi parameter and the two cosmological constants are introduced, and we will comment on the physical relevance of such potentials.

\subsection{The Immirzi parameter and the cosmological constants.}\label{sec:Imm}

One can add the Immirzi parameter $\gamma$ and 2 cosmological constants $\Lambda$ and $\mu$ to (\ref{ModifPleb}) by introducing the following extra terms in the action:\footnote{One can go further and add terms like $\Lambda_1(\phi)\phi^{MN}{}_{IJ}\phi_{MNKL}B^{IJ}\wedge B^{KL}$. Obviously, these can also be accounted for by a field redefinition of $\phi_{IJKL}$. The reason to only discuss the terms in \Ref{ModifPlebConst} is that (i) these are the simplest, and as we will see they already yield interesting phenomenology, and (ii) as these terms have been studied extensively in the non-modified theory, it is important to understand their role in $\Lambda(\phi)$ Plebanski theory.}
\be
S = \int B^{IJ}\wedge F_{IJ} + \gamma^{-1}\star B_{IJ}\wedge F^{IJ} -
 \frac{1}{2}\left(\phi_{IJKL}+\frac{\Lambda+\Lambda(\phi)}{6}\eps_{IJKL}-\frac{1}{3}\mu\delta_{IJKL}\right)B^{IJ}\wedge B^{KL}.\label{ModifPlebConst}
\ee
In the absence of matter, these constants can be absorbed in the definition of $\Lambda(\phi)$, except for the second term if $\gamma = \pm1$ in the Euclidean or $\gamma=\pm i$ in the Lorentzian theory . This is trivial for $\Lambda$, which is absorbed in $\Lambda(\phi)$ in the following, but requires field redefinitions for $\mu$ and $\gamma$. Taking $\hat{\phi}_{IJKL}=\phi_{IJKL}-\f13\mu\d_{IJKL}$ one immediately understands that $\mu$ leads to a theory of the standard form \Ref{ModifPleb}, where in $\Lambda(\phi)$, $\varphi_0$ has to be replaced by $\varphi_0+2\mu$. The discussion of the Immirzi parameter requires a bit more work, which has ben  relegated to appendix \ref{ImmGen}. Apart from showing that for $\g\neq\pm\sqrt{\s}$ the second term in \Ref{ModifPlebConst} can be absorbed in the potential, it is also shown that non-metric gravity is recovered for $\g=\pm\sqrt{\s}$.

Let us discuss the effects of $\g$, $\Lambda$ and $\mu$ on the ST subclass. As one can check, the field redefinitions that were used to demonstrate the above results do not affect \Ref{ScalarPotential}, and accordingly also \Ref{ModifPlebConst} with a potential of the form \Ref{ScalarPotential} and with $\g\neq\pm\sqrt{\s}$ is a ST theory. The different parameters will affect the potential $V(\psi)$, and we will illustrate how these now parameters affect \Ref{PotQuad} and \Ref{LorV}, i.e. the potential obtained from \Ref{LambdaQuad}. The analysis is analogous to that of section \ref{STsubs}: the modified constraints lead to (\ref{BSO4}) or (\ref{BSO31}) and (\ref{TraceInQuad}), and equation \Ref{Bcompat2}, still present since $d_A$ commutes with the hodge dual, can be solved to (\ref{ConnScalTetr}) or (\ref{ConnLor}), which upon substitution in (\ref{ModifPlebConst}) leads to
\bea
S_{\textrm{Eucl}} & = & \int e \left[\tilde{\eta}\frac{1+\g^{-1}-\eta\psi^2(1-\g^{-1})}{2}R-3 \eta (1-\g^{-1})\nabla_{\m}\psi\nabla^{\m}\psi-V_{\text{Eucl}}(\psi)\right],\\ \label{STConstsEucl}
S_{\textrm{Lor}} & = & \int e \left[-\left(\sin(\phi)+\g^{-1}\cos(\phi)\right)R+\f32 \left(\sin(\phi)+\g^{-1}\cos(\phi)\right)\nabla_{\m}\phi\nabla^{\m}\phi - V_{\text{Lor}}(\psi)\right],\label{STConstsLor}
\eea
with
\bea
V_{\text{Eucl}}(\psi) & = & -\frac{1}{a}\frac{(1-\psi^4)^2}{1+\psi^4}+\Lambda(1+\psi^4)+\mu(1-\psi^4),\\
V_{\text{Lor}}(\psi)   & = & \frac{2}{a}\frac{\sin^2(2\phi)}{\cos(2\phi)}+2\Lambda\cos(2\phi)-2\mu\sin(2\phi).
\eea
If $\gamma=1$, $\psi$ is clearly not dynamical in the Euclidean case: (\ref{STConstsEucl}) is just general relativity with an effective cosmological constant (which is of course a member of the non-metric gravity class of theories). To understand the other cases with $\g=\pm\sqrt{\s}$, we make the conformal field redefinition 
$\hat{g}_{\mu\nu}=\tilde{\eta}\frac{1+\g^{-1}-\eta\psi^2(1-\g^{-1})}{2}g_{\mu\nu}$ (again chosing $\eta$ such that the conformal factor is positive), resp. $\hat{g}_{\m\n} = -\left(\sin(\phi)+\g^{-1}\cos(\phi)\right)g_{\m\n}$, in terms of which the action principles are given by
\bea
S_{\textrm{Eucl}} & = & \int e\left[ R - 6\eta \frac{1-\g^{-2}}{\left(1+\g^{-1}-\eta\psi^2(1-\g^{-1})\right)^2}\nabla_{\mu}\psi\nabla^{\mu}\psi
                           -\frac{4 V_{\text{Eucl}}(\psi)}{(1+\g^{-1}-\eta\psi^2(1-\g^{-1}))^2}\right],\label{STImmE}\\
S_{\textrm{Lor}} & = & \int e\left[R - \f32\f{1+\g^{-2}}{\left(\sin(\phi)+\g^{-1}\cos(\phi)\right)^2}\nabla_{\mu}\psi\nabla^{\mu}\psi  
                            - \frac{V_{\text{Lor}}(\psi)}{\left(\sin(\phi)+\g^{-1}\cos(\phi)\right)^2}\right].\label{STImmL}
\eea
It is then clear that any case with $\g=\pm\sqrt{\s}$ simply gives general relativity with an effective cosmological constant. As for $\omega=-\frac{3}{2}$ in Brans-Dicke theory, it is only after going to the Einstein frame that one realizes that the scalar field is not a propagating degree of freedom.

The generic case $\g\neq\pm\sqrt{\s}$ indeed gives a ST theory, for which the ghost properties of the Euclidean theory are unchanged when $\gamma^2>1$: \Ref{STImmE} is stable for $\eta=1$ and unstable for $\eta=-1$. For $\gamma^2<1$ this situation is reversed.\footnote{
This might be surprising at first sight, as we pointed out that for $\gamma\neq\pm 1$ only the potential is affected. One should note however, that $\eta$ is a parameter obtained from $B$, which can be affected by the field redefinition \Ref{Bredef}. This is indeed the case: when B is given by \Ref{BSO4}, 
\beann
\hat{B} & = & \tilde{\eta}\f{1+\g^{-1}-\eta\psi^2(1-\g^{-1})}{2}\star e\wedge e +\tilde{\eta}\f{1+\g^{-1}+\eta\psi^2(1-\g^{-1})}2 e\wedge e\\
             & = & \hat{\tilde{\eta}}\f{1-\hat{\eta}\hat{\psi}^2}2\star \hat{e}\wedge \hat{e} + \hat{\tilde{\eta}} \f{1+\hat{\eta}\hat{\psi}^2}2 \hat{e}\wedge \hat{e},
\eeann
where $\hat{e}=\sqrt{|1+\g^{-1}|}e$, $\hat{\tilde{\eta}}=\tilde{\eta} \textrm{sgn}(1+\g^{-1})$, $\hat{\psi} = \sqrt{\left|\f{1-\g^{-1}}{1+\g^{-1}} \right|}$ and most importantly $\hat{\eta}=\eta \textrm{sgn}(1-\g^{-2})$. It is this last $\hat{\eta}$ which determines the stability. The correct interpretation is not that $\g$ affects the stability (there is still a stable and an unstable sector), but that $\g$ affects the value of $\eta$, which labels the sectors.
} As before, the Lorentzian theory is ghost-free for any real value of the Immirzi parameter.

We focus on the stable sectors, for which the field transformation needed to obtain a canonical scalar kinetic term is given by 
\bea
\mathfrak{so}(4) &:&
\hat{\psi}(\psi) = 
\left\{
\begin{array}{lr}
\sqrt{6}\arctanh(\sqrt{\eta\frac{1-\g^{-1}}{1+\g^{-1}}}\psi),                           & \psi^2\eta\frac{1-\g^{-1}}{1+\g^{-1}}<1; \\
\sqrt{6}\arctanh(\sqrt{\eta\frac{1+\g^{-1}}{1-\g^{-1}}}\psi^{-1}),                   & \psi^2\eta\frac{1-\g^{-1}}{1+\g^{-1}}>1;
\end{array}\right.\\
\mathfrak{so}(3,1) &:&
\psi(\phi) = 
\left\{
\begin{array}{lr}
\sqrt{6}\arctanh(\f{1-\g^{-1}\tan(\f\phi2)}{\sqrt{1+\g^{-2}}}),                           &\phi \in ]2\arctan(\g-\sqrt{1+\g^2}),2\arctan(\g+\sqrt{1+\g^2})[; \\
\sqrt{6}\arctanh(\f{\sqrt{1+\g^{-2}}}{1-\g^{-1}\tan(\f\phi2)}),                     &\phi \in ]2\arctan(\g+\sqrt{1+\g^2}),2\arctan(\g-\sqrt{1+\g^2})[ .
\end{array}\right.\label{TrafoSO31}
\eea
Let us discuss the potential for the Euclidean theory first. In the first sector, the potential is, immediately dropping the hats, given by
\be\label{EuclPot}
V = 
\f4{(1+\g^{-1})^2} \cosh^4(\f\psi{\sqrt{6}})\left( -\f1a\f{\left(1-\b^2\tanh^4(\f\psi{\sqrt{6}})\right)^2}{1+\b^2\tanh^4(\f\psi{\sqrt{6}} )} +\m+\Lambda +(\Lambda-\mu)\b^2\tanh^4(\f\psi{\sqrt{6}})\right) 
\ee
where $\beta=\f{1+\g^{-1}}{1-\g^{-1}}$. The potential in the second sector can be obtained from \Ref{EuclPot} by replacing $\g\rightarrow -\g$ and $\m\rightarrow-\m$.
For small $\psi$, these potentials behave as a free field,
\be
V(\psi)\approx \f4{(1+\g^{-1})^2}\left(\Lambda+\mu-a^{-1}\right) +\f43\f 1{(1+\g^{-1})^2}\left(2\Lambda + 2\mu - a^{-1}\right)\psi^2,\quad |\psi|\ll1, 
\ee
while for large fields they behave like
\be
V(\psi)\approx 
\left\{
\begin{array}{lr}
\f1{(1-\g^{-2})^2}\left(-\f{\g^{-2}}{8a(1+\g^{-2})}+\f\Lambda2(1+\g^{-1})-\m\g^{-1}\right)e^{\f{4}{\sqrt{6}}|\psi|}, & \g^{-1}\neq 0\; {\textrm{or}}\; \Lambda\neq0\\
2\mu e^{\f2{\sqrt{6}}|\psi|}, & \g^{-1}=0,\Lambda = 0, \mu\neq 0,\\
-\f8a, &  \g^{-1}=0,\Lambda=0, \mu = 0
\end{array}\right., \quad |\psi|\gg 1.
\ee
The effect of $\Lambda$, $\g$ and $\m$ is to cause an exponential behavior at large values of the fields. As has already been noted in the introduction, such potentials have attracted interest in cosmology, due to the energy density which tracks the matter/radiation energy density at early times, when $|\psi|\gg1$,
and their behaviour as pressureless cold dark matter at later times when $|\psi|\ll1$ \cite{Matos2000,Sahni2000}. Scalar fields with such a potential are referred to as scalar field dark matter (SFDM) in the following.

For the Lorentzian case, the explicit inverse of \Ref{TrafoSO31} with $\g^{-1}\neq 0$ is more complicated. As the explicit formula for the general potential is long and not very instructive, we discuss the effects of the cosmological constants and the Immirzi parameters separately. For $\g^{-1}=0$, the potential is given by
\be
V = \f8a\f{\sinh^2(\sqrt{\f23}\psi)}{\sinh^2(\sqrt{\f23}\psi)-1} - 4\m\sinh(\sqrt{\f23}\psi) - 2\Lambda \left[\sinh^2(\sqrt{\f23}\psi)-1\right],\label{VLor1}
\ee
the second sector being related to the first one by replacing $\m\rightarrow -\m$. As in the case $\Lambda=\m=0$, this potential has 2 singularities, at $|\sinh(\sqrt{\f23}\psi)|=1$, which separate 3 sectors. When $\Lambda=\m = 0$ we found that only the bounded sector had a minimum, for negative $a$. While \Ref{VLor1} displays the same behavior in the bounded sector, it is changed in the unbouded sectors $|\sinh(\sqrt{\f23}\psi)|>1$. A minimum is found in both of these sectors iff $\Lambda<0$, $a>0$ (and for such parameters, both sectors are stable). For $\Lambda=0$, $\m\neq0$, only the sector with $-\textrm{sgn}(\m)\sinh(\sqrt{\f23}\psi)>1$ is stable (and displays a minimum). The behavior of the potential for large values is given by
\be
V(\psi)\approx 
\left\{
\begin{array}{lr}
-2\Lambda e^{\f4{\sqrt{6}}|\psi|}, & \Lambda\neq 0\\
-4\m e^{\f2{\sqrt{6}}\psi},                & \Lambda=0,\m\neq 0\\
\f8a						  & \Lambda=0=\m
\end{array}\right., \quad |\psi|\gg 1.
\label{Lorlimits}
\ee
While it is clear that, for $|\psi|$ taking initial value $\gg 1$, the singularities in the potential will not affect the interpretation as SFDM in the unbounded sectors with mimimum, the effect of these singularities for generic initial values is still unclear.


On the other hand, setting $\m=\Lambda=0$, the potential in the first sector reduces to
\be
V= \f{32}{a(1+\g^{-2})^2}\f{\left(\g+\g^{-1}+2\g\cosh^2(\f\psi{\sqrt{6}})(\sqrt{1+\g^{-2}}\tanh(\f\psi{\sqrt{6}})-1)\right)^2}{\g^2\left(\f{1+\g^{-2}}{\cosh^2(\f\psi{\sqrt{6}})(1-\sqrt{1+\g^{-2}}\tanh(\f\psi{\sqrt{6}}))}-2\right)^2-4}.\label{LorVImm}
\ee
As in the previous cases, there are two singularities for $\g^2\neq1$, but at different values of the field, given by 
\be
\psi = \sqrt{6}\arctanh\left(\f{1-\g^{-1}\left(\pm1\pm\sqrt{2}\right)}{\sqrt{1+\g^{-2}}}\right),\label{Immsings}
\ee
where among the 4 choices of signs, only the 2 for which the argument of $\arctanh$ is in $]-1,1[$ are selected. For $\g^2=1$, only one singularity arises, at $\psi_{|1|}=\sqrt{6}\arctanh(\f{2-\sqrt{2}}{\sqrt{2}})$, and hence the theory is only split in 2 sectors. The behavior at $\pm\infty$ is given by
\be
V  \approx 
\left\{
\begin{array}{lr}
\f{2}{a(1-\g^{-2})} \left(\f{1\mp\sqrt{1+\g^{-2}}}{1+\g^{-2}}\right)^2e^{\f{4|\psi|}{\sqrt{6}}}, & \g^2\neq 1,\\
-\f{(1\mp \sqrt{2})^3}{8a}e^{\sqrt{6}|\psi|}						              , & \g^2=1.
\end{array}\right.,
\ee
the $\mp$ signs reflecting the fact that the coefficients at $\pm\infty$ differ. In the unbounded sectors, the potential for $\g^2\neq1$ has a minimum and the theory is stable iff $a(1-\g^{-2})>0$. When $\g^2=1$, there is always 1 stable and 1 unstable sector: for $a>0$, the sector $]\psi_{|1|},\infty[$ is stable, while for $a<0$ the sector $]-\infty,\psi_{|1|}[$ is. It is noteworthy that in the limit $\g\rightarrow \pm\infty$, the potential \Ref{LorVImm} does not approach \Ref{LorV}. Both singularities \Ref{Immsings} tend to $+\infty$ and the behavior at $\pm\infty$ will in this limit be given by
\be
V\approx 
\left\{
\begin{array}{lr}
\f8a e^{-4\f\psi{\sqrt{6}}},                 & \psi\rightarrow -\infty\\
\f{\g^{-4}}{2a}e^{4\f\psi{\sqrt{6}}},   &\psi \rightarrow +\infty 
\end{array}\right..
\ee
The fact that the limit $\g^{-1}\rightarrow 0$ does not repreduce the theory without Immirzi parameter, is due to the singularity of the field transformation \Ref{TrafoSO31} in this limit. The behaviour of the potential in the second sector is analogous.

The above analysis reveals the differences between the 2 reality conditions that can be imposed on the Lorentzian theory, referred to at the end of section \ref{sec:PlebSec}. If the tetrads $e^{(+)}$ and $e^{(-)}$ in \Ref{BDecomp} are required to be real, it is straightforward to adopt the analysis of the Euclidean case discussed above. One finds $i$ times \Ref{SFinal}, and consequently the theory is unstable in the sector in which $\eta=-\bar{\eta}$. If on the other hand, all fundamental fields, i.e. $B$, $A$ and $\phi$, are taken to be real, we found that the theory is stable in all sectors. The viability of the theory with real fundamental fields eliminates the need for reality conditions which impose reality of the tetrads, and from this point on, we restrict the discussion to the Lorentzian theory with all fundamental fields real.


\section{Matter Coupling} \label{sec:MC}

In order to interpret the results of the previous section, matter has to be coupled to the theory, i.e. we need to study which matter couplings terms ${\mathcal L}_m$ in \Ref{ST} can arise from \Ref{ModifPleb} with matter coupled to it. The precise form of ${\mathcal L}_m$ will significantly affect the phenomenology of \Ref{ST}. For example, with a generic coupling ${\mathcal L}_m(Q,g_{\m\n},\phi)$ the weak equivalence principle (WEP) will be violated, which would be manifested by the presence of a fifth force. The WEP will hold for test particles, i.e. neglecting self-interaction, when all matter fields couple to the same metric $\hat{g}_{\m\n}=A^2(\phi)g_{\m\n}$, i.e. when the coupling term takes the form
\be
S_m(A^2(\phi)g_{\m\n},Q). \label{STmetriccoupl}
\ee
All test particles then move on the geodesics of $\tilde{g}_{\m\n}=A^2(\phi)g_{\m\n}$, the matter coupling singling out this conformal frame as the physical one. It is however useful to point out that it was shown in \cite{Damour1994} that the metric coupling \Ref{STmetriccoupl} is an attractor of a more general class of coupling terms, in which the masses of the different particles are taken to be a function of the scalar fields (with the extrema of these different mass functions coinciding). The deviations from \Ref{STmetriccoupl} at the present epoch would be very small in such a scenario, but cannot a priori be excluded at early times. 


What kind of matter coupling can be expected to arise from \Ref{ModifPleb} with matter coupled to it? There are different ways to address this question. In \cite{KrasnovSymmBr}, a symmetry breaking mechanism was proposed, in which a vacuum solution of \Ref{ModifPleb} with an extended gauge group $G$ will break the gauge symmetry down to $SU(2)$ times its centralizer. As such, a unified description of gravity and matter fields is offered by $B$, with the $SU(2)$ part describing the graviton, the centralizer part describing the matter fields. It is however still unclear how the standard model fermionic fields can be recovered in this framework, and we will not pursue this approach here. However, the results of the previous sector are a special case of such a unified description, both the gravitational and the scalar field being contained in $B$, in which the coupling between both fields is understood in detail. The scalar field is minimally coupled to $-\sin(\phi)g_{\m\n}$, which is the imaginary part of the Urbantke metric. While it is not clear how to generalize this result to the different matter fields\footnote{In general it is not clear whether the imaginary part of the Urbantke metric is Lorentzian. It is possible to construct two independent real Lorentzian metrics from real $B^{IJ}_{\m\n}$, taking the real and imaginary part of $e^{(+)}$ as fundamental tetrads, but in the ST subclass, using these for parametrizing $B$ has not proven very useful.}, it can serve as a guiding principle for doing so.

A less ambitious approach is to couple matter fields `by hand', i.e. by adding a term 
\be
S(B,Q).\label{MatCoupl}
\ee 
This approach of coupling by hand is more general than the unified approach, which will amount to a specific form of \Ref{MatCoupl}.  Note that fermionic matter fields will require a term $S(B,A,\zeta)$, which does not pose extra complications. Elegant coupling terms have been proposed to couple matter to the Plebanski formalism in this way. Most notably scalar and Yang-Mills type matter can be coupled polynomially to $B$ directly, as opposed to constructing the non-polynomial Urbantke metric and using the standard minimal coupling action \cite{CapovillaDellJacobsonMason}. Such Ans\"atze can be consider also in the modified theory, but in this section we study the general form \Ref{MatCoupl}, as was done for non-metric gravity in \cite{KrasnovGeod}. If one is to understand \Ref{ModifPleb} as an effective quantum gravity action, the only restrictions on the matter Lagrangian are invariance under gauge transformations and diffeomorphism, and the recovery of minimally coupled matter in the Plebanski limit $\Lambda(\phi)\rightarrow \Lambda$. The questions we set out to answer are (i) what conditions should, in the general class of $\Lambda(\phi)$ theories, be imposed on \Ref{MatCoupl} to obtain a viable theory, (ii) what is the phenomenology of couplings satisfying these conditions in the ST subclass, more in particular for which $S(B,Q)$ is the scalar field minimally coupled to the metric wrt which matter couples (this is the coupling needed to derive the SFDM phenomenology referred to above)?


Before proceeding, note that the two Urbantke metrics $g^{(+)}_{\m\n}=e^{(+), I}_\m e^{(+), J}_\n\eta_{IJ}$ and $g^{(-)}_{\m\n}=e^{(-), I}_\m e^{(-), J}_\n\eta_{IJ}$ are complex for real $B$. There are many ways of defining a pair of real metrics. While the explicit formula in terms of $B$ is less compact than proposals used in previous literature\footnote{
Compact explicit formula are easily obtained for the real and imaginary parts of $-g^{(+)}g^{(+),\m\n}={\mathpzc g}^{(1)}{\mathpzc g}^{(1),\m\n}+i{\mathpzc g}^{(2)}{\mathpzc g}^{(2),\m\n}$ \cite{Capovilla2001,Beke2012} or of $\sqrt{-g^{(+)}}g^{(+)}_{\m\n}$:
\bea
\sqrt{- g^{(r)}}g^{(r)}_{\m\n} & =& \f1{6}\eta_{IN}\d_{JMKL} B^{IJ}_{\m\a}\tilde{B}^{KL,\a\b}B^{MN}_{\b\n},\\
\sqrt{-g^{(i)}}g^{(i)}_{\m\n} & =& \f1{12}\eta_{IN}\eps_{JMKL} B^{IJ}_{\m\a}\tilde{B}^{KL,\a\b}B^{MN}_{\b\n},
\eea
with $\tilde{B}^{IJ,\m\n}=\f12\eps^{\m\n\r\s}B^{IJ}_{\r\s}$. For solutions (\ref{BSO31}) of the relaxed constraints, one can check by direct calculation, or by using $g^{(+)}_{\m\n}=(\cos(\phi)-i\sin(\phi))g_{\m\n}$, that
\bea
{\mathpzc g}^{(1)}{\mathpzc g}^{(1),\m\n} =   \f{1}{8}\cos(\phi)(\cos^2(\phi) - 3\sin^2(\phi)) g g^{\m\n}, &&\quad {\mathpzc g}^{(2)}{\mathpzc g}^{(2),\m\n}  = \f18\sin(\phi) (\sin^2(\phi)-3\cos^2(\phi))g g^{\m\n}\\
\sqrt{-g^{(r)}}g^{(r)}_{\m\n}  = -\sin(\phi) (\sin^2(\phi)-3\cos^2(\phi))\sqrt{-g}g_{\m\n},&&\quad \sqrt{-g^{(i)}}g^{(i)}_{\m\n}  = \cos(\phi)(\cos^2(\phi) - 3\sin^2(\phi))\sqrt{-g}g_{\m\n}.
\eea
}, the real and imaginary part of the Urbantke metric, which for \Ref{BSO31} are given respectively by
\be
g^{(+)}_{\m\n}= g^{(2)}_{\m\n}+ig^{(1)}_{\m\n}=\cos(\phi)g_{\m\n}-i\sin(\phi)g_{\m\n}, \label{UrbRI}
\ee
will be more useful in the following. Note that in the presence of the scalar constraint, these reduce to
\bea
\text{top}: && g^{(1)}_{\m\n} =0,\;\;\;\;\;\; \quad g^{(2)}_{\m\n} =  \pm g_{\m\n}\\
\text{gr}: &&  g^{(1)}_{\m\n} = \pm g_{\m\n}, \quad g^{(2)}_{\m\n} = 0.
\eea

Consider the most general matter term
\be
S(B,Q) = S[g^{(1)}_{\m\n},g^{(2)}_{\m\n}, b^{(1)}_{ij},b^{(2)}_{ij},Q],\label{TensorMatter}
\ee
with $b^{(1)}, b^{(2)}$ the real and imaginary part of $b^{(+)}$. Under the condition
\be
{\mathcal L}(g_{\m\n},0,\d^i_j,\d^i_j,Q) ={\mathcal L}_m(g_{\m\n},Q),\label{LimitCond}
\ee
with ${\mathcal L}_m$ the minimally coupled matter Lagrangian, a sector with matter minimally coupled to general relativity will be found in the `Plebanski limit' $\Lambda(\phi) \rightarrow \Lambda$.\footnote{More precisely we consider a sequence of potentials for which both $\Lambda(\phi)\rightarrow \Lambda$ and $\f{\partial \Lambda}{\partial \phi^{IJKL}} \rightarrow 0$, the second condition being a consequence of the first one for polynomial invariants.} This is easily understood: in the Plebanski limit, the simplicity constraints, and consequently a sector in which $B= \star e\wedge e$ is a solution, are recovered. Substituting this solution in the action, one finds with \Ref{LimitCond} that matter is minimally coupled to the Einstein-Cartan action. Many Ans\"atze satisfy this criterium, and in the unmodified case all are equivalent.

As we assume that the Lagrange multipliers are decoupled from the matter fields, the (modified) simplicity constraints are not affected by matter. Hence, in the ST subclass, we are interested in 
\be
S[-\sin(\phi)g_{\m\n},\cos(\phi)g_{\m\n}, \d_{ij},\d_{ij},Q],
\ee
which is almost a generic ST matter coupling term $S[g_{\m\n},\phi,Q]$ where it not for the condition $\displaystyle \lim_{\phi\rightarrow-\f\pi2}S[g_{\m\n},\phi,Q]=S_m[g_{\m\n},Q]$. Generically, this dependence affects the signs of the kinetic terms of the matter fields, implying a more complicated splitting in stable and unstable sectors. The following coupling reduces to \Ref{STmetriccoupl} in the sector $\sin(\phi)<0$:
\be
S[g^{(1)}_{\m\n},g^{(2)}_{\m\n}, b^{(1)}_{ij},b^{(2)}_{ij},Q] = S_m[A^2(g^{(2),\m\n}g^{(1)}_{\m\n})g^{(1)}_{\m\n},Q]. 
\ee
Two types of coupling terms satisfying \Ref{LimitCond} are however of particular interest. 

As discussed in \cite{Tennie2010}, an example of a class of matter Lagrangians yielding minimal coupling is
\bea
{\mathcal L}(g^{(1)}_{\m\n},g^{(2)}_{\m\n}, b^{(1)}_{ij},b^{(2)}_{ij},\phi_{IJKL},Q) & = & \f12 {\mathcal L}_m[-ig^{U+},Q]+\frac{1}{2}{\mathcal L}_m[ig^{U-},Q]\label{TennieCoupling}\\
 & = &  \f12 {\mathcal L}_m[g^{(1)}-ig^{(2)},Q]+\frac{1}{2}{\mathcal L}_m[g^{(1)}+ig^{(2)},Q]\label{TennieCoupling2}
\eea
Such a coupling obviously satisfies \Ref{LimitCond}, and once the (A)SD split is made, this is the most natural real coupling term doing so. In the modified theory however, such a coupling involves some subtleties with respect to its stability. In the ST subclass, it reduces to 
\be
\f12\mathcal{L}_m[e^{i\left(\phi+\f\pi2\right)}g_{\m\n},Q]+\f12\mathcal{L}_m[e^{-i\left(\phi+\f\pi2\right)}g_{\m\n},Q].\label{TennieST}
\ee
For a massive scalar field $\Phi$, the matter Lagrangian $\mathcal{L}_m=-\sqrt{-g}\left(\nabla_\m\Phi\nabla^\m\Phi+m^2\Phi^2\right)$ coupling to \Ref{LorSTJordan}, is then
\be
\int \sqrt{-g}\left[\sin(\phi)\nabla_\m\Phi\nabla^\m\Phi -m^2\left(\sin^2(\phi)-\cos^2(\phi))\Phi^2\right)\right]\label{TennieScal}
\ee
The signs of the kinetic terms are all equal, in both sectors $\sin(\phi)<0$ and $\sin(\phi)>0$, and hence do not incite instabilities, which can be demonstrated most clearly by going to the Einstein frame.\footnote{With our sign convention $(-+++)$, one usually takes the signs in front of $\nabla_\m\Phi\nabla^\m\Phi$ to be negative, such as to have a positive contribution to the kinetic energy. This is of course only a convention, as the classical theory is not affected by multiplying the action by a global minus sign. When the potential terms give a contribution to the action which is bounded from below, a theory with all kinetic terms negative, such as \Ref{LorSTJordan}+\Ref{TennieScal} in the sector $\sin(\phi)>0$, is stable.} 
The potential terms however are more subtle. Note that, due to the singularity in \Ref{ModifScalLor} the theory will, in addition to the splitting $\sin(\phi)<0$, $\sin(\phi)>0$, always be split in two sectors subject to $|\cos(\phi)|>|\sin(\phi)|$ resp. $|\cos(\phi)|>|\sin(\phi)|$. We thus find 2 stable sectors: one with positive kinetic and negative potential contributions to the action ($\sin(\phi)<0$ and $|\sin(\phi)|>|\cos(\phi)|$), and one with negative kinetic and positive potential contributions to the action ($\sin(\phi)>0$ and $|\cos(\phi)|>|\sin(\phi)|$). In the other sectors, the theory is unstable.
For a Yang-Mills field, \Ref{TennieST} yields, both in the Einstein and the Jordan frame, a coupling term
\be
\int -\f14\sqrt{-g}g^{\m\a}g^{\n\b}F_{\m\n}F_{\a\b}.
\ee
Hence, contrary to the conclusions for the scalar field, the sign of the kinetic terms is only correct in the sector $\sin(\phi)<0$. 

While \Ref{TennieCoupling} does yield stable sectors, deciding on which sector is stable has to be done a posteriori. We therefor propose to consider the simpler
\be
{\mathcal L}(g^{(1)}_{\m\n},g^{(2)}_{\m\n}, b^{(1)}_{ij},b^{(2)}_{ij},\phi_{IJKL},Q) = {\mathcal L}_m(g^{(1)}_{\m\n},Q),\label{NaturalCoupling}
\ee
which also satisfies \Ref{LimitCond}.
Matter then couples to the Jordan frame action \Ref{LorSTJordan} by ${\mathcal L}_m(-\sin(\phi)g_{\m\n},Q)$. Hence, in the sector $\sin(\phi)<0$, it couples to the Einstein frame action by ${\mathcal L}_m(g_{\m\n},Q)$. In the absence of the Immirzi parameter, matter is then minimally coupled and in this sector no additional stability issues arise. Note furthermore that this is, in the absence of the Immirzi parameter but in the presence of the cosmological constants, the coupling needed to derive the SFDM phenomenology referred to above. 




\section{BF formulation of scalar-tensor theories}\label{sec:JBD}
It is immediate from the above discussion that a massless ST theory can be obtained by relaxing the scalar component of the constrains, without introducing a potential. In other words, the action
\be
S_{\text{ST-BF}}[B^{IJ},A^{IJ},\lambda_{IJKL}]=\int B^{IJ}\wedge F_{IJ}-\frac{1}{2}\lambda_{IJKL}B^{IJ}\wedge B^{KL},
\ee
with $\lambda\in \mathbf{(2,0)\oplus(0,2)\oplus (1,1)}$, i.e. $\lambda_{IJKL}=\lambda_{KLIJ}=-\lambda_{JIKL}$ and $\delta^{IJKL}\mu_{IJKL}=\eps^{IJKL}\mu_{IJKL}=0$, is a BF formulation of massless ST theories. It is not difficult to show this: the constraints 
\be
B^{IJ}\wedge B^{KL}=-\f1{12}\eps^{IJKL}<B\wedge\star B> + \f16\d^{IJKL}<B\wedge B>
\ee
are solved by (\ref{BSO31}) and the resulting compatibility equation $d_AB=0$ is solved  by (\ref{ConnScalTetr}). Substituting these solutions in the action, it reduces the massless ST theory
\be
\int \sqrt{-g}\left[-\sin(\phi)R+\f32\sin(\phi)\nabla_{\m}\phi\nabla^{\m}\phi\right],\label{MasslST}
\ee
where the matter coupling will differentiate between the different massless theories. As an illustration, we will derive the matter coupling corresponding to JBD theory. As the kinetic terms will always have the right sign, only the stable theories $\omega>-\f32$ can be recovered.
Taking matter to be coupled by a term $S_m[\tilde{g}_{\m\n},Q]$, with $\tilde{g}_{\mu\nu}=\Omega(\phi)^{-2}g_{\mu\nu}$, one can make a conformal field redefinition $\hat{g}_{\m\n}=\tilde{g}_{\m\n}$, to obtain an action principle, with the hats immediately dropped,
\be
\int e\Omega^4\left[-\sin(\phi)\Omega^{-2}R+6\sin(\phi)\Omega^{-3}\nabla_{\mu}\nabla^{\mu}\Omega+\f32\Omega^{-2}\sin(\phi)\nabla_\mu\phi\nabla^\mu\phi\right] + S_m[g_{\m\n},Q].
\ee
In order to compare this action to the JBD action, 
\be
\int e\left[\psi R-\frac{\omega}{\psi}\nabla^{\mu}\psi\nabla_{\mu}\psi\right] + S_m[g_{\m\n},Q],
\ee
one should make a field redefinition $\psi(\phi)=-\sin(\phi)\Omega^2$, to find that these actions are equivalent up to boundary terms iff 
\be
\frac{\omega}{\sin(\phi)\Omega^2}\nabla_\mu\left(\sin(\phi)\Omega^2\right)\nabla^\mu\left(\sin(\phi)\Omega^2\right)=-6\nabla_\mu\left[\Omega\sin(\phi)\right]\nabla^\mu\Omega+\f32\sin(\phi)\Omega^{2}\nabla_\mu\phi\nabla^\mu\phi.
\ee
This results in the differential equation
\be
(\omega+\f32)\sin(\phi)\left(\frac{d\Omega}{d\phi}\right)^2+(\omega+\f32)\cos(\phi)\Omega\frac{d\Omega}{d\phi}+\left(\f\omega{\sin(\phi)}-(\omega+\f32)\sin(\phi)\right)\Omega^2=0,
\ee
which has no non-zero solution for the unstable theory $\omega<-\f32$ or the 2 DOF theory $\omega=-\f32$, but which splits in two first order differential equations 
\be
\f{d\Omega}{d\phi}=\f12\left(-\cot(\phi)\pm\f{\sqrt{\f{3/2}{\omega+3/2}}}{\sin(\phi)}\right)\Omega
\ee
for $\omega>-\f32$. These are solved by
\be
\Omega_{\pm}^2(\phi) = (|\sin(\phi)|)^{-1\pm\sqrt{\f{3/2}{\omega+3/2}}}(1+\cos(\phi))^{\mp\sqrt{\f{3/2}{\omega+3/2}}} ,
\ee
which means that to recover the JBD theory, matter has to be coupled to \Ref{MasslST} by either $S_m[\Omega_+^{-2}g_{\m\n},Q]$ or $S_m[\Omega_-^{-2}g_{\m\n},Q]$. Note that, just like \Ref{SFinalLor} and JBD theory, \Ref{MasslST} describes two sectors: a sector in which $\sin(\phi)>0$, i.e. $\psi(\phi)<0$ and the physical sector (in which the signs of the ST and matter kinetic terms are equal) with $\sin(\phi)<0$ and consequently $\psi(\phi)>0$. In the latter, by using the formula
\be
\tan(\phi)=-\f1{12}\f{\left<B\wedge B\right>}{\sqrt{-\det g^{(2)}}},
\ee
$\cos(\phi)$ and $\sin(\phi)$ can easily be constructed from the bivectors.\footnote{By using $\cos(\phi)=-\f{\text{sgn}(\tan(\phi))}{\sqrt{1+\tan^2(\phi)}}$ and $\sin(\phi) = -\frac{|\tan(\phi)|}{\sqrt{1+\tan^2(\phi)}}$.}
In conclusion, the action
\begin{subequations}
\begin{gather}
S = S_{\text{ST-BF}}[B,A,\mu] + S_m\left[A^2_\omega\left(B\right)g^{(1)}_{\m\n},Q\right]\label{JBDMatt},\\
A^2_\omega(B)  = \left(\frac{\left|\f{\left<B\wedge B\right>}{\sqrt{-g^{(2)}}}\right|}{\sqrt{1+\left(\f{\left<B\wedge B\right>}{\sqrt{-g^{(2)}}}\right)^2}-\text{sgn}\left(\f{\left<B\wedge B\right>}{\sqrt{-g^{(2)}}}\right)}\right)^{\pm\sqrt{\f{3/2}{\omega+3/2}}},
\end{gather}
\end{subequations}
constitutes a BF formulation of JBD theory.

While JBD theory can also be obtained by simply coupling the scalar fields along the lines of section \ref{sec:MC}, here we exploited the fact that this scalar DOF is present already in the bivector field $B$. In conclusion, we have shown that JBD theory can be described as a constrained BF theory, but also for the massless case it is more natural to consider different ST theories.


\section{Conclusion}

The 4-dimensional low energy string effective action of the massless modes is a scalar-tensor theory coupled to various gauge fields and fermions. We showed that also in $\Lambda(\phi)$ Plebanski theory, which has been advocated as an effective quantum gravity action, a scalar-tensor theory arises. We isolated the subclass in which only these scalar-tensor modes are propagating, and the scalar mode was identified as the conformal factor relating the two Urbantke metrics that can be constructed from the $\mathfrak{so}(3,1)$-valued bivectors. This subclass was then shown to be equivalent to the Bergmann-Wagoner-Nordvedt class of scalar-tensor theories, with a massive scalar particle. This is the main difference with the string effective action, which a priori is invariant under dilatations (though it is not clear how compactification affects this invariance). 

We analyzed the consequences of different reality conditions, and showed that the simplest choice, requiring the bivectors to be real, leads to a viable theory. At least in this scalar-tensor subclass, there is no need to impose reality of the two Urbantke metrics. Working under the condition of a real bivector field, the scalar field is minimally coupled to the metric $g^{(1)}$, which is the imaginary part of the Urbantke metric. If the matter fields are also taken to be minimally coupled to $g^{(1)}$, the scalar field decouples. Whether such a coupling scheme arises naturally from a unified gravity-matter description for non-scalar matter fields is subject to further research.

Finally, we showed that the scalar-tensor theories which arise from a quadratic $\Lambda(\varphi_0)$ with Immirzi parameter or cosmological constant, yield dark matter phenomenology. This incites the following question, which we do not attempt to answer here. In \cite{Drummond2001}, it was shown that in bi-metric theories of gravity, the conformal degree of freedom causes an exponentially fading repulsion, modifying Newton's law at large distances in such a way that the galactic rotation curves can be accounted for. This is closely related to the dark matter phenomenology found in this paper, as the scalar mode is precisely the conformal factor relating the 2 Urbantke metrics. However, in standard bi-gravity theories this conformal degree of freedom corresponds to a ghost, which propagates with the wrong sign of the kinetic term, implying its repulsive character.
Due to the enhanced gauge symmetry of the potentially ghost-free bi-metric theories of gravity recently proposed in \cite{Rosen}, this unstable mode can be gauge fixed. As the remarks above, and the analysis of this paper, suggest that it is precisely this scalar mode which is responsible for the dark matter phenomenology, it is important to understand whether the results of \cite{Drummond2001} extend to these ghost-free theories.

\begin{acknowledgments}
The author is pleased to thank Simone Speziale for many stimulating discussions on $\Lambda(\phi)$ Plebanski theory, Norbert Van den Bergh for reading the manuscript, and Kirill Krasnov for asking about the coupling between the scalar and the gravitational field. The author is supported by the Research Foundation-Flanders (FWO), and wishes to thank the
Centre de Physique Th«{e}orique, of the Aix-Marseille University for its hospitality during a research visit.
\end{acknowledgments}

\appendix


\section{Notation}\label{App:Not}
We are concerned with $\mathfrak{so}(4)$ and $\mathfrak{so}(3,1)$, which are represented by 4x4 matrices $X^I{}_J$ anti-symmetric, i.e. $X^{IJ}=-X^{JI}$, when the indices are raised by  $\eta^{IJ}=(\sigma,+,+,+)$, with $\s=1$ for $\mathfrak{so}(4)$ and $\s=-1$ for $\mathfrak{so}(3,1)$. One can define 2 different metrics on this space: 
\benn
\d_{IJKL}=\eta_{I[K}\eta_{L]J}
\eenn
and the totally anti-symmetric $\eps_{IJKL}$ normalized such that $\eps^{0123}=1$. $\eps$ can be used to define the Lie algebra Hodge dual
\be
\star X^{IJ}=\f12\eps^{IJ}_{KL}X^{KL}.
\ee
$X$ can be decomposed in its selfdual part (which is an eigenvector of $\star$ with eigenvalue $\s\sqrt{\s}$) and its anti-selfdual (ASD) part (having eigenvalue $-\s\sqrt{\s}$), and the projectors on the (A)SD part are given by
\be
P^{IJ}_{(\eps)}{}_{KL}=\f12\left(\d^{IJ}{}_{KL}+\f{\eps\sqrt{\sigma}}2\eps^{IJ}{}_{KL}\right),
\ee
$\eps=+$ denoting the SD part and $\eps=-$ denoting the ASD part:
\be
\star P^{IJ}_{(\eps)}{}_{KL}=\eps\s\sqrt{\s} P^{IJ}_{(\eps)}{}_{KL}.
\ee
$\mathfrak{su}(2)$ labels can be introduced on the (A)SD subspace by choosing a (in the Lorentzian case timelike) direction and taking
\benn
P^{IJ}_{(\eps),i}=2P^{IJ}_{(\eps)}{}_{0i}=\d^{IJ}_{0i} + \f{\eps\sqrt{\sigma}}2\eps^{IJ}{}_{0i},
\eenn
normalized such that
\benn
P^{IJ}_{(\eps),i}P^{KL}_{(\eps),i} = \s P_{(\eps)}^{IJKL}, \quad \d_{IJKL}P^{IJ}_{(\eps),i}P^{KL}_{(\eps),j}=\sigma\d_{ij}.
\eenn
Given a tetrad, one can construct $\mathfrak{su}(2)$-valued two-forms by mapping the algebra indices to spacetime indices:
\benn
\Sigma^i_{(\eps)}[e]=P^i_{(\eps),IJ}e^I\wedge e^J.
\eenn
These bivectors are (A)SD with respect to the spacetime hodge dual, and satisfy the following useful identity
\benn
\Sigma^i_{\eps}[e]\wedge\Sigma^j_{(\eps)}[e]=2\sqrt{\s}\eps e\d^{ij}.
\eenn

We will regularly consider fields $\phi^{IJKL}$ taking value in the symmetric direct product of the Lie algebras, i.e. $\phi^{IJKL}=\phi^{KLIJ}=-\phi^{JIKL}$. We will denote such symmetrization by adding a line, e.g. $\phi^{IJKL}=\phi^{(IJ|KL)}=\f12\phi^{IJKL}+\f12\phi^{KLIJ}$. Most important for this paper is the case in which $\eps_{IJKL}\phi^{IJKL}=0$, such that $\phi$ takes value in $\bf{(0,2)\oplus(2,0)\oplus(1,1)\oplus (0,0)}$:
\beann
\phi^{IJKL} & = & \left(P^{IJ}_{(+)}{}_{MN}P^{KL}_{(+)}{}_{PQ}-\f13P^{IJKL}_{(+)}P_{(+)MNPQ}\right)\phi^{MNPQ} + \left(P^{IJ}_{(-)}{}_{MN}P^{KL}_{(-)}{}_{PQ}-\f13P^{IJKL}_{(-)}P_{(-)MNPQ}\right)\phi^{MNPQ} \\
                    &&\quad + 2 P^{(IJ}_{(+)}{}_{MN}P^{KL)}_{(-)}{}_{PQ}\phi^{MNPQ} + \f16\d^{IJKL}\d_{MNPQ}\phi^{MNPQ}\\
 & = & \left(P^{IJ}_{(+),i}P^{KL}_{(+),i}-\f13\d_{ij}P^{IJ}_{(+),k}P^{KL}_{(+),k}\right) \phi^{(+)}_{ij} +  \left(P^{IJ}_{(-),i}P^{KL}_{(-),i}-\f13\d_{ij}P^{IJ}_{(-),k}P^{KL}_{(-),k}\right) \phi^{(-)}_{ij} \\
                     &&\quad +   P^{(IJ}_{(+),i}P^{KL)}_{(-),j}\psi_{ij} + \f16\d^{IJKL}\varphi_0\\
 & \equiv & {\bf \Pi}^{IJKL}_{{\bf(2,0)},ij}\phi^{(+)}_{ij} + {\bf \Pi}^{IJKL}_{{\bf(0,2)},ij}\phi^{(-)}_{ij} + {\bf \Pi}^{IJKL}_{{\bf(1,1)},ij}\psi_{ij} + 
{\bf \Pi}^{IJKL}_{{\bf(0,0)}} \varphi_0,
\eeann
where we have introduced a suitable parametrization for the irreducible components
\benn
\phi^{(\eps)}_{ij}= \left(P^{IJ}_{(\eps),i}P^{KL}_{(\eps),j}-\f13\d_{ij}(P^{IJ}_{(\eps),k}P^{KL}_{(\eps),k}\right)\phi_{IJKL}, \quad \psi_{ij}= 2P^{IJ}_{(+),i}P^{KL}_{(-),j}\phi_{IJKL}, \quad \varphi_0 = \d_{IJKL}\phi^{IJKL}.
\eenn

Finally, we use angled brackets to denote taking the inner product with respect to $\d$, e.g. for the Lorentz-algebra valued bivectors $B$
\be
\left\langle B\wedge B\right\rangle = \d_{IJKL} B^{IJ}\wedge B^{KL}, \quad \left\langle B\wedge\star B\right\rangle = \f12\eps_{IJKL} B^{IJ}\wedge B^{KL}.
\ee

\section{General effects of the Immirzi parameter} \label{ImmGen}
In this appendix we discuss the effects of introducing the Immirzi parameter to $\Lambda(\phi)$ Plebanski gravity, in the absence of matter:
\be
S = \int B^{IJ}\wedge F_{IJ} + \gamma^{-1}\star B_{IJ}\wedge F^{IJ} -
 \frac{1}{2}\left(\phi_{IJKL}+\frac{\Lambda(\phi)}{6}\eps_{IJKL}\right)B^{IJ}\wedge B^{KL}.\label{ModifPlebImm}
\ee
Because of the phenomenologically interesting effects in the scalar-tensor subsector, an explicit analysis for $\Lambda(\phi)=\Lambda_0(\varphi_0)$ is discussed in the main text. The analysis for general potentials $\Lambda(\phi)$ is given here.
For $\g\neq\pm\sqrt{\sigma}$, \Ref{ModifPlebImm} is equivalent to $\Lambda(\phi)$ Plebanski gravity without the Holst term, but with a different potential $\tilde{\Lambda}_{\g}(\phi)$. This mechanism is responsible for the dependence of the mass of the second graviton on the Immirzi parameter \cite{Beke2012}. When $\g = \pm\sqrt{\sigma}$, the theory reduces to the non-metric theory of gravity, which is the $\mathfrak{su}(2)$ analog of $\Lambda(\phi)$ Plebanski gravity.

\subsection{Generic case: $\g\neq\pm\sqrt{\s}$}
In order to eliminate the Holst term $\gamma^{-1}\star B_{IJ}\wedge F^{IJ}$, a number of field redefinitions have to made. First, take 
\be
\hat{B}^{IJ}=B^{IJ}+\gamma^{-1}\star B^{IJ}, \label{Bredef}
\ee
which can be inverted to $B^{IJ}=\frac{1}{1-\s\g^{-2}}(\hat{B}^{IJ}-\g^{-1}\star \hat{B}^{IJ})$, in terms of which
\be
S=\int\hat{B}^{IJ}\wedge F_{IJ} - \f1{2(1-\s\g^{-2})^2}\left(\tilde{\phi}_{IJKL}+\f16 \left[\Lambda(\phi)(1+\s\g^{-2})-\g^{-1}\varphi_0\right]\eps_{IJKL} \right)\hat{B}^{IJ}\wedge\hat{B}^{KL},\label{SImmTemp}
\ee
where 
\be
\tilde{\phi}_{IJKL}=\phi_{IJKL} - 2\g^{-1}{}^*\phi_{(IJ|KL)}+\g^{-2}{}^*\phi^*_{MNPQ} + \f16 \g^{-1}\varphi_0\eps_{IJKL} -\f23\s\g^{-1}\Lambda(\phi)\d_{IJKL},\label{phitild}
\ee
with $^*\phi_{IJKL}=\f12\eps_{IJ}^{MN}\phi_{MNKL}$, $\phi^*_{IJKL}=\f12\eps_{KL}^{MN}\phi_{IJMN}$.
There are 2 classes of solutions to the equations of this theory. In the first class 
\be
\f{\partial \Lambda(\phi)}{\partial \varphi_0}\neq \f14\left(\s\g+\g^{-1}\right),
\ee
and one can take \Ref{phitild} as a field redefinition, since in this class \Ref{phitild} can be inverted to $\phi_{IJKL}(\tilde{\phi}_{IJKL})$. In that case, \Ref{SImmTemp} reduces to the familiar form
\be
S = \int \hat{B}^{IJ}\wedge F_{IJ} -\f12\left[\hat{\phi}_{IJKL}+\f16\tilde{\Lambda}_\g(\hat{\phi})\eps_{IJKL}\right]\hat{B}^{IJ}\wedge\hat{B}^{KL},
\ee
where a rescaling $\hat{\phi}=\f1{(1-\s\g^{-2})^2}\tilde{\phi}$ has been made, and where the precise form of the potential depends on the explicit form of the inverse $\phi(\hat{\phi})$:
\be
\tilde{\Lambda}_\g(\hat{\phi}) = \Lambda(\phi(\hat{\phi}))\left(1+\s\g^{-2}\right)-\g^{-1}\varphi_0(\hat{\phi}).
\ee

The second class will only be present when $\f{\partial \Lambda(\phi)}{\partial \varphi} = \f14\left(\s\g+\g^{-1}\right)$ has solutions, and this condition will be identically satisfied in this class. Varying \Ref{SImmTemp} wrt $\varphi$, one obtains $\left\langle \hat{B}\wedge\star \hat{B}\right\rangle = 0$. Hence the variation of the action wrt $\phi_{IJKL}$ is equivalent to
\be
B^{IJ}\wedge B^{KL} = \f16\d^{IJKL}\left\langle B\wedge B\right\rangle.
\ee
In order to solve these equations, one can make an irreducible decomposition as in \Ref{SimplConsIrr}, the only difference being that the scalar component of the constraints gives $e^{(+)}m^{(+)}+e^{(-)}m^{(-)}=0$. Hence $m^{(+)ij}=m^{(-)ij}=\d^{ij}$ and $e^{(-)I}_\m=\psi e^{(+)I}_\m$, and the conformal factor $\psi$ is determined by 
\be
e^{(-)}=-e^{(+)}.
\ee
In the $\mathfrak{so}(4)$ case, no non-degenerate real solutions exist. In the $\mathfrak{so}(3,1)$ case, one finds four solutions
\be
B^{IJ}=\pm\f1{\sqrt{2}}e^I\wedge e^J \pm \f1{\sqrt{2}} (\star e\wedge e)^{IJ}. 
\ee
Substituting this in the action, one obtains plus or minus the Holst action, with $\g=\pm1$. 

\subsection{Non-metric case: $\g=\pm\sqrt{\s}$}
For this case, it is convenient to rewrite the action in terms of the irreducible components of $B$ and $\phi$, which for $\g=\sqrt{\s}$ yields
\bea
S & = & \int B^i_{(+)}\wedge F^i_{(+)}-\f12\left(\phi^{(+)}_{ij}+\tilde{\Lambda}\d_{ij}\right)B^i_{(+)}\wedge B^j_{(+)}-\f12\psi_{ij}B^i_{(+)}\wedge B^j_{(-)}-\f12\Phi_{ij}B^i_{(-)}\wedge B^j_{(-)} ,\label{Sg+}\\
\Phi_{ij} & = & \phi^{(-)}_{ij}+(\f\s6\varphi_0-\f{\s\sqrt{\s}}3\Lambda)\d_{ij}, \\
\tilde{\Lambda} & = & \f{\s\varphi_0}6+\f{\s\sqrt{\s}\Lambda}3.
\eea
$B^i_{-}$ is no longer a dynamical field, and we are only concerned with the dynamics of $B^i_{(+)}$ and $A^i_{(+)}$.  It is convenient to make the field redefinition $(\phi^{(-)}_{ij},\varphi_0)\rightarrow \Phi_{ij}$ which is invertible whenever $\sqrt{\s}\f{\partial\Lambda}{\partial \varphi_0}\neq \f12$. Varying \Ref{Sg+} wrt $\varphi_0$, one finds that $B^i_{(+)}\wedge B^i_{(+)}=0$ in the sector with $\sqrt{\s}\f{\partial\Lambda}{\partial \varphi_0} = \f12$. It then follows from the variation wrt $\phi^{(+)}$ that $B^i_{(+)}\wedge B^j_{(+)}=0$, and this sector reduces to the degenerate $\mathfrak{su}(2)$ Plebanski sector. Henceforth we focus on the sector with $B^i_{(+)}\wedge B^i_{(+)}\neq 0$.

Varying \Ref{Sg+} wrt  $B^i_{(+)}$ and $A^i_{(+)}$, gives the respective equations
\bea
F^i_{(+)}         & = & \left(\phi^{(+)}_{ij}+\tilde{\Lambda}\d_{ij}\right)B^i_{(+)} + \f12\psi_{ij}B^j_{(-)},\label{FBpl}\\
d_{A_{(+)}}B & = & 0.\label{Apl}
\eea
The other equations are algebraic, and are given by
\bea
B^i_{(+)}\wedge B^j_{(+)}-\f13\d^{ij}B^k_{(+)}\wedge B^k_{(+)} & = & -\f{\partial \tilde{\Lambda}}{\partial \phi^{(+)}_{ij}}B^k_{(+)}\wedge B^{(+)},\label{modifmetr}\\
B^i_{(-)}\wedge B^j_{(-)} & = &-\f{\partial \tilde{\Lambda}}{\partial \Phi^{ij}}B^k_{(+)}\wedge B^k_{(+)},\label{BminASD}\\
B^i_{(-)}\wedge B^j_{(+)} & = &-\f12\f{\partial\tilde{\Lambda}}{\partial\psi_{ij}} B^k_{(+)}\wedge B^k_{(+)},\label{BminSD}\\
\Phi_{ij}B^i_{(-)}+\psi_{ij}B^i_{(+)} & = & 0\label{Phipsi}.
\eea
We are interested in the case with $B_{(+)}^i\wedge B_{(+)}^j\equiv h^{ij}$ an invertible matrix: in that case (up to a conformal factor) a metric can be constructed from the bivectors, by requiring that $B_{(+)}^i$ is selfdual with respect to the space-time hodge dual. The $B_{(+)}^i$'s then form a basis for the space of selfdual 2-forms, and $B^i_{(-)}$ can be decomposed in its space-time SD and ASD parts,
\be
B^i_{(-)}=Q^i_jB^j_{(+)}+\tilde{B}^i_{(-)}.
\ee
Equation \Ref{BminSD} can be solved for the coefficients $Q^i_j$ of the self-dual part, and equation \Ref{BminASD} then determines 
\be
\tilde{B}_{(-)}^i\wedge \tilde{B}_{(-)}^j= -\f14\f{\partial\tilde{\Lambda}}{\partial \psi_{ki}}h^{-1}_{kl}\f{\partial\tilde{\Lambda}}{\partial \psi_{lj}}h^2-\f{\partial \tilde{\Lambda}}{\partial\Phi_{ij}}h.\label{BASDwedge}
\ee 
Substituting this decomposition in equation \Ref{Phipsi}, one finds 
\be
\psi_{ij} = -Q^k_i\Phi_{kj}, \quad \Phi_{ij}\tilde{B}^j_{(-)} = 0.
\ee 
Substituting the solution\footnote{Note that in general, $\Phi$ is constrained to be a zero divisor of \Ref{BASDwedge}. Also for possible non-zero solutions of these equations one can show that $\psi_{ij}B^{j}_{(-)}=0$. Either this equation together with \Ref{modifmetr} can be solved for $\Phi$ and $\phi^{(+)}$ in terms of $h^{ij}$, or these equations give constraints on $h^{ij}$. The former case gives exactly the non-metric theories after integrating out the Lagrange multiplier $\phi_{ij}$. The latter case appears to give a possible generalization of the non-metric theories, still with 2 DOFs but with a constrained auxiliary field (e.g. $\tr (h^2)=c$).} $\Phi_{ij}=\psi_{ij}=0$ in equations \Ref{FBpl}, \Ref{Apl}, \Ref{modifmetr} one finds exactly the field equations of non-metric gravity, with the action principle given by
\be
\int B_{(+)}^i\wedge F^i(A_{(+)}^i)-\f12\left(\phi^{(+)}_{ij}+\f{\s\sqrt{\s}}{3}\Lambda(\phi^{(+)}_{ij}P^{IJ}_{(+),i}P^{KL}_{(+),j})\d_{ij}\right)B_{(+)}^i\wedge B_{(+)}^j.
\ee

\end{document}